\documentclass[amsmath,amssymb,aps,showpacs,floatfix,pre,twocolumn ]{revtex4-1}

\usepackage{graphicx}
\usepackage{bm}
\begin{document}

\title{Mode-coupling approach for the slow dynamics of a liquid on a spherical substrate}
\author{Julien-Piera Vest,  Gilles Tarjus and Pascal Viot}
\affiliation{$^1$Laboratoire de Physique Th\'eorique de la Mati\`ere Condens\'ee, CNRS  UMR 7600, UPMC-Sorbonne Universit{\'e}s,
 4, place Jussieu, 75252 Paris Cedex 05, France}

\begin{abstract}
We study the dynamics of a one-component liquid constrained on a spherical substrate, a 2-sphere, and investigate how the mode-coupling theory (MCT) can 
describe the new features brought by the presence of curvature. To this end we have derived the MCT 
equations in a spherical geometry. We find that, as seen from the MCT, the slow dynamics of liquids in curved space at low 
temperature does not qualitatively differ from that of glass-forming liquids in Euclidean space. 
The MCT predicts the right trend for the evolution of the relaxation slowdown with curvature but is dramatically off at a quantitative level.
\end{abstract}

\newcommand{\thedate}{\today}
\maketitle

\section{Introduction}

What is the influence of the curvature of space on the dynamics of a fluid? This question arises in several physical situations, such as colloidal assemblies on droplets in particle-stabilized (Pickering) emulsions 
\cite{Aveyard2003,Tarimala2004,Bausch2003,Einert2005,Lipowsky2005,Irvine2010,Leunissen20022007,Irvine2012} or fluid monolayers
adsorbed on  large solid particles \cite{Post1986,Post1988}. According to the frustration-based theory of glass-forming liquids \cite{Nelson2002,Tarjus2005}, 
depending on the dimension of space and the nature of the liquid, curvature may induce or release ``geometric frustration'' 
and therefore affect the ability to form glassy phases and the slowdown of relaxation. For spherical particles in 2 dimensions, hexagonal or hexatic $6$-fold order 
is prevalent in flat (Euclidean) space, and a constant nonzero curvature thwarts the long-range or quasi-long-range ordering  
by forcing in an irreducible density of topological defects \cite{PhysRevB.62.8738,Nelson2002,Tarjus2005,Bowick2007,Sausset2008,Sausset2010,Sausset2010a,Tarjus2011}.
At high density and/or low temperature the particle configurations then consist of a hexatic background interrupted 
by topological defects in the form of finite-length strings known as ``grain boundary scars'' \cite{PhysRevB.62.8738,Bausch2003,Einert2005,Lipowsky2005,Bowick2007,Sausset2010,Irvine2010}. 
Correspondingly, there is a strong slowing down of the dynamics as one lowers the temperature or increases the density. One may wonder if this dynamical slowdown is similar or not to that of conventional glass-forming liquids?

Here we address these questions through a study of the dynamics of a one-component atomic liquid embedded in a 2-sphere, \textit{i.e.}, a surface of constant positive 
curvature, by Molecular Dynamics simulation and the mode-coupling theory (MCT).
The MCT of glass formation was developed in the mid-80s in the context of the theory of simple liquids \cite{Bengtzelius1984,Gotze2008}. 
In its canonical version the MCT predicts the dynamics of a system from the mere knowledge of the static pair correlation function or static structure factor. 
It is a mean-field-like theory of the glass transition which predicts a dynamical, ergodic to nonergodic, transition that is not observed in real glass-formers in finite dimensions where it is (at best) replaced by a crossover.  
Due to the absence of singularity in finite dimensions, the predictions are not quantitatively accurate ($T_c$ is typically overestimated by a factor of $2$ 
in simple glass-forming systems). Nonetheless, the MCT appears to correctly capture the observed trends of the dynamics in physical situations where a significant change 
in the static structure factor is encountered, as, \textit{e.g.}, in attractive colloids and other soft-matter systems\cite{Letz2000,Pham2002,Henrich2007,Berthier2010}.

The points we have therefore investigated within a generalization of the MCT for a spherical geometry are the following:

1) Is the predicted dynamical transition governed by a different type of singularity than that found for standard glass-forming liquids? For instance, it has been predicted that glass-forming liquids confined in disordered porous media may display within the MCT, depending on the model parameters, a continuous (A-type)  transition different from the discontinuous (B-type) one observed in bulk glass-formers \cite{Krakoviack2005}. Is a similar phenomenon encountered for particles on a sphere due to specificities of the kinetics of the grain boundary scars moving in a random and slower evolving hexatic environment?

2) Is the MCT able to predict the variation of the slowing down of relaxation with curvature? To make the comparison more quantitative, we have carried out Molecular Dynamics simulations of a model one-component liquid on $S^2$ with several values of the curvature, parametrized by the ratio of the radius of the embedding sphere over the particle diameter, and we have fitted the numerical data according to MCT predictions.

We find that the MCT singularity on $S^2$  is actually of the same type as that in Euclidean space. 
Thus, as seen from the MCT, there is nothing special about glass formation in curved space, which make liquids in curved space \textit{bona fide} models for studying the generic features 
of the glass transition. As for the answer to the second question, we conclude that the MCT predicts the right trend,  with the the slowdown becoming more marked as curvature is reduced. However, the effect is 
found to be much too weak when compared to the simulation data. This is another example \cite{Berthier2010b} where the MCT is unable to properly amplify the small changes seen in the static pair structure to 
describe the observed strong differences in dynamics.

\section{A simple glass-forming liquid on a sphere}

\subsection{Model and simulation}
We consider a monodisperse fluid of $N$ atoms interacting through a pairwise Lennard-Jones potential and embedded either in a $2$-sphere (the surface of a 3-dimensional ``ball'') or in the Euclidean plane $E^2$.
The pair interaction potential is given by
\begin{equation}
\label{interactionLJ}
 v( r)=4\epsilon\left[\left(\frac{\sigma}{r}\right)^{12}-\left(\frac{\sigma}{r}\right)^6\right],
\end{equation}
where $r$ is the geodesic distance between two atom centers.  The interaction is truncated at a conventional cutoff 
distance of $2.5 \sigma$. On a 2-sphere $S^2$ ($S^2$ usually refers to the 2-sphere of radius unity but we extend the notation to a sphere of any radius $R$), this interaction is also known as the 
``curved line of force'' \cite{Post1986}. The units of 
mass, length, energy and time are $m$, $\sigma$, $\epsilon$, and $\sqrt{m\sigma^2/\epsilon}$. On the surface of a sphere of radius $R$, the reduced density is fixed by the number of particles and the ratio $R/\sigma$, 
which represents a dimensionless radius of curvature, as
\begin{equation}
\label{eq_reduced_density}
\tilde \rho=\frac{2N}{\pi}\left(1-\cos\left(\frac{\sigma}{2R}\right)\right)\,.
\end{equation}
In the Euclidean (flat) limit, $R\rightarrow \infty$, one then recovers the conventional reduced density, $\tilde \rho\rightarrow (N/V)\sigma^2$, where $V$ is the volume (actually, the area) of the system.

For studying the dynamics of the fluid on the sphere it is convenient to view each particle as a 3-dimensional rotator rigidly linked to the center of the sphere so that it is constrained to move at a fixed distance $R$ of this center. The Hamiltonian is then 
\begin{equation}
 H=\sum_{i=1}^N \frac{m{\boldsymbol \omega}_i^2}{2mR^2}-\frac{1}{2}\sum_{i\neq j}{v(r_{ij})}\,,
\end{equation}
where ${\boldsymbol \omega}_i$ is the angular velocity of the rotator $i$, and the associated equations of motion read 
\begin{equation}
 m\dot{{\boldsymbol \omega}}_i=-{\bf r}_i\times \sum_{j\neq i}\partial_ {\bf r_i} v(r_{ij})\,,
\end{equation}
with $r_{ij}$ the geodesic distance between rotators $i$ and $j$.

A Molecular Dynamics simulation can then be implemented via a ``velocity Verlet algorithm'', with an update of the angular velocities for a time step $\Delta t$ according to \cite{Vest2014}
\begin{equation}
\begin{aligned}
{\boldsymbol \omega}_i (t+\Delta t)=& {\boldsymbol \omega}_i (t)  
 -\frac{\Delta t}{2m}\big[{\bf r}_i (t+\Delta t) \times \sum_{j\neq i}
 \partial_ {\bf r_i} v(r_{ij})(t+\Delta t) \\&
 +{\bf r}_i (t)\times \sum_{j\neq i} \partial_ {\bf r_i} v(r_{ij})(t)\big]
\end{aligned}
\end{equation}
and, by using ${{\bf v}_i} =\boldsymbol{\omega}_i \times {\bf r}_i $, an update of the  positions  as
\begin{equation}
 {\bf r}_i (t+\Delta t)= [1+a(t,\Delta t)] {\bf r}_i (t)+ \Delta t\big [{\boldsymbol \omega}_i (t)+\frac{\Delta t}{2}\dot{{\boldsymbol \omega}}_i\big ]\times {\bf r}_i,
\end{equation}
where $a(t,\Delta t)$ is determined by enforcing  the constraints  ${\bf r}_i^2 (t+\Delta t)={\bf r}_i^2 (t)=R^2$ \cite{Ryckaert1977,Lee_Leok_2009}. See Ref. [\onlinecite{Vest2014}] for more details.

For illustration we have studied systems at a reduced density $\tilde \rho=0.92$ for a range of temperature $T$ and of ratio $R/\sigma$. For the spherical geometry, initial configurations consist of  particles placed 
randomly on $S^2$ such that  the distance between any pair of particles is always larger than $0.85\sigma$. In the first stage of the simulation, 
the velocities are often rescaled in order that the mean kinetic energy becomes equal to the chosen temperature $T$. For the Euclidean geometry, we have used a rectangular cell of  aspect ratio equal to 
$2/ \sqrt{3}$ with periodic boundary conditions to minimize the number of topological defects. The initial configuration is then taken as a triangular lattice. As for $S^2$, 
the velocities are rescaled in the early stage of the simulation in order to enforce that the system stays at a given temperature $T$. 

\subsection{Structure and dynamics}
We have characterized the structure and the dynamics of the system by computing two quantities, the static structure factor $S(k)$ and the self-intermediate scattering function $F_s(k,t)$. 
(Other observables are considered and described in a previous paper \cite{Vest2014}.) 
In $S^2$, the static structure factor is expressed as (see also the Appendix \ref{Fourier_S2} for details)
\begin{align}
\label{eq_structure_factor}
S(k)&=\frac{1}{N}\sum_{i,j} \bigg<P_k\bigg(\cos\bigg(\frac{r_{ij}}{R}\bigg)\bigg)\bigg>,
\end{align}
where $k$ is an integer, $P_{k}$ is the $k$th Legendre polynomial, and $r_{ij}$ is the geodesic distance between atoms $i$ and $j$. The self-intermediate scattering function is given by \cite{Tarjus2011}
\begin{align}
F_s(k,t)&=\frac{1}{N}\sum_{j} \bigg<P_k\bigg(\cos\bigg(\frac{r_{j}(0,t)}{R}\bigg)\bigg)\bigg>,
\end{align}
where $r_{j}(0,t)$ is the geodesic distance traveled by atom $j$ between times $0$ and $t$.

The static structure factor is the necessary input for the MCT equations. We have obtained it from Molecular Dynamics simulation for different  system sizes, 
$N=1000$, $2000$, $4000$, and $12000$, on $S^2$. For the fixed chosen reduced density $\tilde \rho=0.92$ these sizes correspond to an increasing reduced radius of 
curvature,  $R/\sigma \simeq9.3$, $13.2$, $18.5$ and $32.3$. Note that the chosen system sizes, with $N\geq 1000$ are large enough to avoid trivial finite-size effects 
that are unrelated to curvature and would be already present in the Euclidean plane. 
In addition we have also studied the one-component liquid on the Euclidean plane, above and near the ordering transition. 
In this case we have used systems that are sufficiently large, up to $14400$, so that the static pair correlation function reaches its asymptotic value 
for a distance less than half the linear size of the simulation cell at the temperature studied.

\begin{figure}[t]
\begin{center}   
  \includegraphics[width=0.45\textwidth]{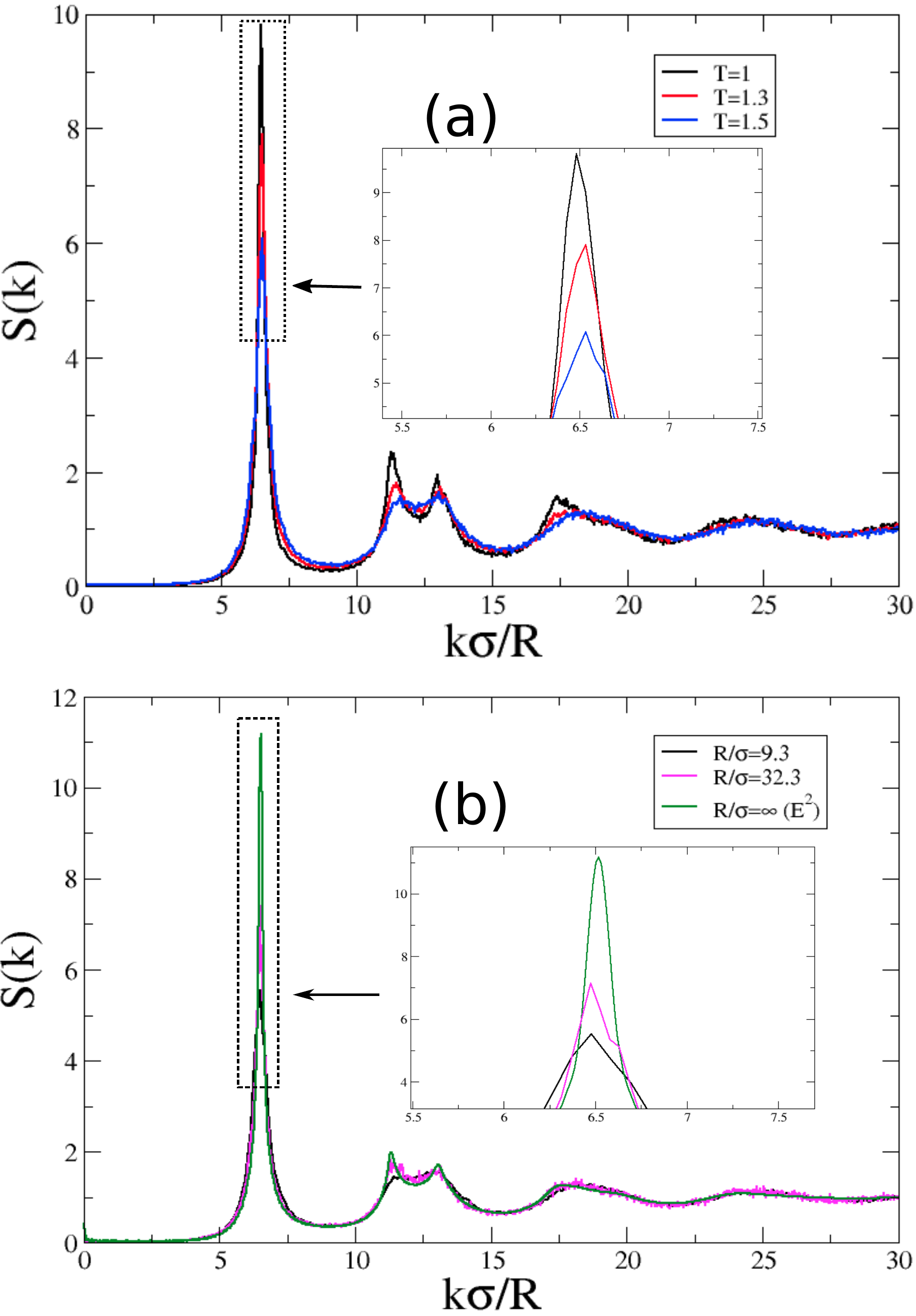}
   \caption{Static structure factors $S(k)$ of the one-component Lennard-Jones liquid in
   2-dimensional space versus the reduced wavenumber $k\sigma/R$. 
   (a) Effect of temperature at a constant curvature $R/\sigma=18.5$: $T=1$, $1.3$ and $1.5$. (b) Effect of curvature at constant temperature $T=1.45$: $R/\sigma=9.3$, $R/\sigma=32.3$ and Euclidean case (in this case the data is plotted versus $k\sigma$ where $k$ is the wavevector modulus). In all cases the reduced density is $\tilde \rho=0.92$. The insets show a zoom in on the first peak.}
   \label{Sk_Tcomp}
\end{center}
\end{figure}

In Fig. \ref{Sk_Tcomp}, we display the effect of the temperature at constant curvature (a) and the effect of curvature at constant temperature (b) on the static 
structure factor. As can be seen, the variation is modest and unremarkable in all cases. The main variation appears through the value of the first peak, especially
when varying curvature at fixed temperature. Correspondingly, a change in the (split) second peak is also observed.

\begin{figure}[t]
\begin{center}   
\includegraphics[width=0.45\textwidth]{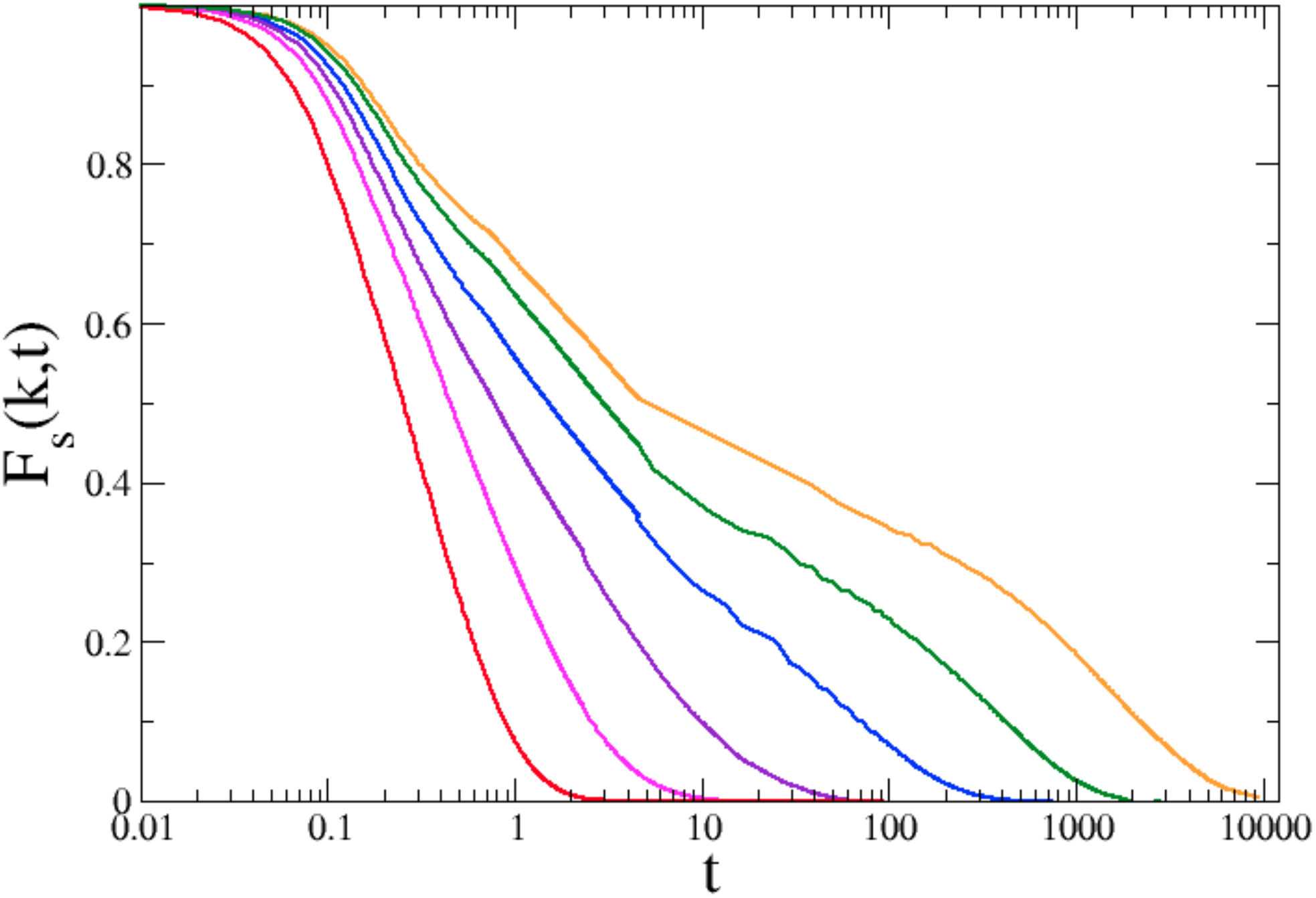}
   \caption{Self-intermediate scattering function $F_s(k,t)$ versus time on a log-linear plot for $R/\sigma= 18.5$. The temperature varies from $T=3$ (left curve) down to $T=0.6$ (right curve).}
   \label{Intermediate_scattering}
\end{center}
\end{figure}

\begin{figure}[h!]
\begin{center}
\includegraphics[width=0.45\textwidth]{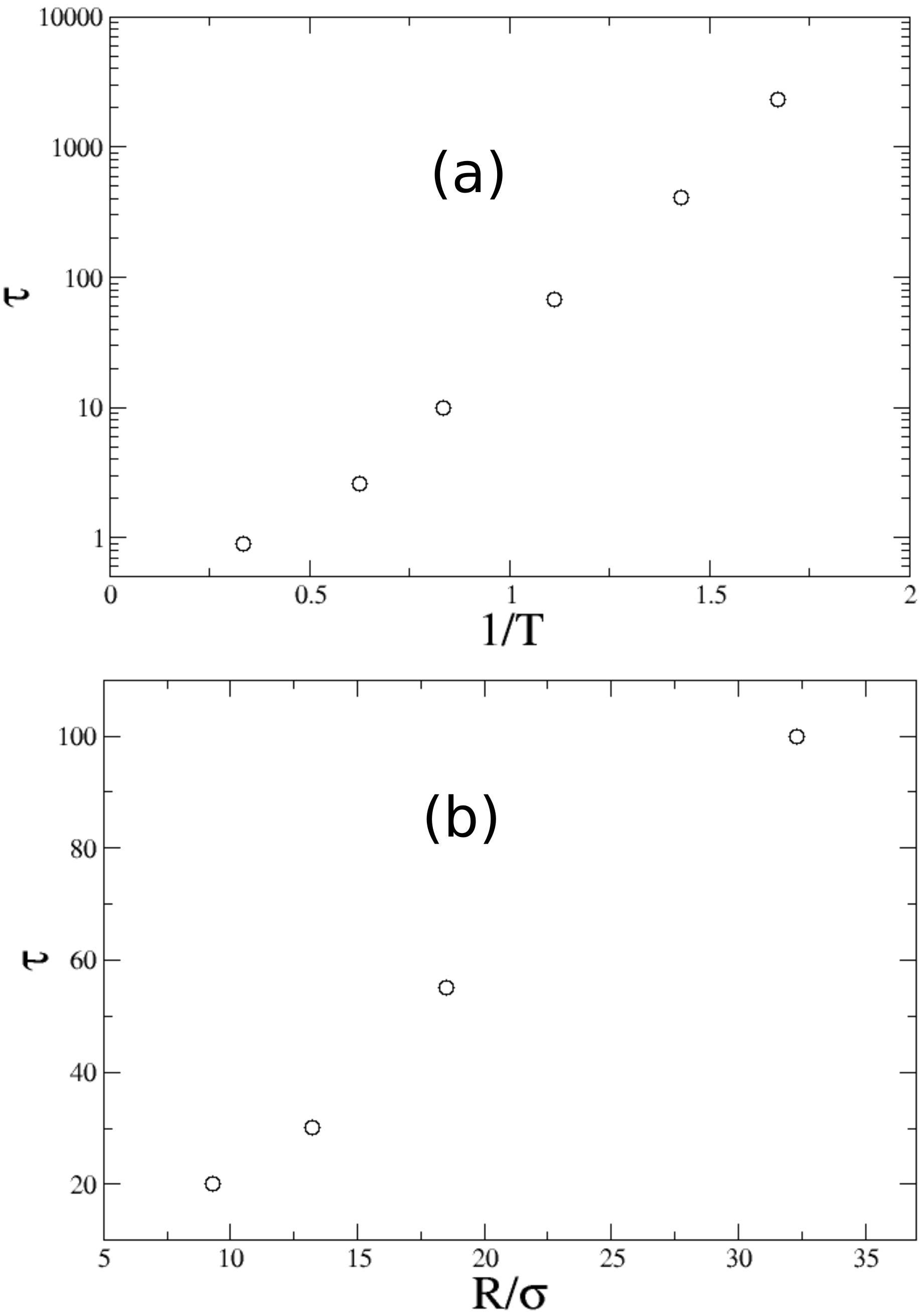}   
   \caption{(a) Arrhenius plot of the relaxation time $\tau$ versus $1/T$ for $R/\sigma= 18.5$. (b) Variation with curvature of the relaxation time for a fixed temperature $T=0.9$. Note the difference in scale for the $y$-axis in the two plots.}
   \label{relax_time}
\end{center}
\end{figure}

We illustrate in Fig. \ref{Intermediate_scattering} the evolution with temperature of the self-intermediate scattering function $F_s(k,t)$, for $k$ corresponding to the typical
interatomic distance, in the case 
$R/\sigma= 18.5$. With decreasing temperature, one can see the gradual change from a simple exponential decay to a more stretched, two-step-like, decay. However, in the range of temperature under study, 
there is no well-defined plateau in the scattering function.   Fig. \ref{relax_time}  shows the relaxation time  $\tau$ as a function of $1/T$ 
for a given curvature, $R/\sigma= 18.5$, 
and as a function of curvature for a given temperature, $T=0.9$ ($\tau$ is defined from $F_s(k,\tau)=0.1$). As expected, curvature thwarts crystallization and the systems remains liquid down to temperatures below the ordering transition in the Euclidean plane 
(which is around $1.3-1.4$). One also finds, as already observed for a liquid in a surface of constant negative curvature \cite{Sausset2008,Sausset2010,Sausset2010a}, 
that the relaxation time increases with decreasing curvature, \textit{i.e.} with increasing $R/\sigma$.

\section{Mode Coupling Theory in spherical geometry}

\subsection{Preliminaries}

To derive the generalization of the MCT equations for a liquid embedded in a $2$-sphere, we repeat the steps of the derivation in Euclidean space and introduce the appropriate changes associated with the positive curvature.

For a monodisperse system on $S^2$, the microscopic density $\rho\big(\Omega)$ is defined  as
\begin{align}
\rho\big(\Omega=(\theta,\phi)\big)&=\sum_{i=1}^N \frac{1}{R^2}\delta(\Omega-\Omega_i),\nonumber\\
&=\sum_{i=1}^N \frac{1}{R^2\sin\theta_i}\delta(\theta-\theta_i)\delta(\phi-\phi_i)\,,
\end{align}
where $\theta$ and $\phi$ are the azimuthal and polar angles, respectively. By using Eq. (\ref{eq:dft}), the Fourier transform $\rho_{k,l}$ of the density reads 
\begin{equation}
\rho_{k,l}=\sqrt{\frac{4\pi}{2k+1}}\sum_{i=1}^N{Y_{k,l}^*(\theta_i,\phi_i)}\,,
\end{equation}
where $k$ and $l$ are  integers such that $k\geq 0$ and  $-k\leq l\leq k$, and $Y_{k,l}^*$ is (the complex conjugate of) a spherical harmonics (see Appendix \ref{Fourier_S2}).
The dynamic structure factor is defined from the density fluctuations $\delta \rho=\rho-\langle\rho\rangle$ as 
\begin{equation}
 F(k,t)=\frac{1}{N}\sum_{l=- k}^k\big<\delta \rho_{k,l}^*(0)\delta \rho_{k,l}(t)\big>,
\end{equation}
where the bracket denotes the thermal average.
Because of the homogeneity of the system, $\big<\delta \rho_{k,l}^*(0)\delta \rho_{k,l}(t)\big>=\big<\delta \rho_{k,0}^*(0)\delta \rho_{k,0}(t)\big>$ and
\begin{equation}
 F(k,t)=\frac{2k+1}{N}\big<\delta \rho_{k,0}^*(0)\delta \rho_{k,0}(t)\big>.
\end{equation}
At time $t=0$ the above function reduces to the static structure factor, $F(k,t=0)=S(k)=4\pi/N\sum_{ij}\langle Y_{k,0}(\Omega_i)Y_{k,0}(\Omega_j)\rangle$ [which coincides with Eq. (\ref{eq_structure_factor})].

The MCT starts with a generalized Langevin equation for the Fourier components of the density and their time derivative.
One first  defines  the vector $ A_{k,l}=\left(  \begin{array}{c}\delta \rho_{k,l}  \\-i\delta \dot{\rho}_{k,l} \end{array}  \right)$ and the associated correlation matrix $\mathbf{C}_{k,l}(t)=\big<A_{k,l}^*(0)A_{k,l}(t)^\top\big>$. 
Because of the space homogeneity,   $\mathbf C_{k,l}(t)=\mathbf C_{k,0}(t)\equiv \mathbf C_k(t)$, and
\begin{eqnarray}
\mathbf C_k(t)=\left(  \begin{array}{cc}
\frac{NF(k,t)}{2k+1} & \frac{N}{2k+1}\dot F(k,t)  \\
-\frac{N}{2k+1}\dot F(k,t) & \big<\delta \dot{\rho}_{k,0}^*(0)\delta \dot{\rho}_{k,0}(t)\big> \end{array}  \right)
\end{eqnarray}
where a dot indicates a time derivative.

By using the projection-operator technique,
the evolution equation for the correlation matrix can be cast in the form of a generalized Langevin equation \cite{Reichman2005},
from which it is easy to derive the following equation for the normalized dynamic structure factor $f(k,t)=F(k,t)/S(k)$:
\begin{align}
&\ddot f(k,t)+\frac{4\pi k_B T}{m V}\frac{k(k+1)}{S(k)}f(k,t) + \nonumber\\ 
&\int_{0}^t{dt' K(k,t-t')\dot f(k,t')}=0 
\label{exact}
\end{align}
where $V=4\pi R^2$ is the surface area of the sphere, the memory function $K(k,t)$ is defined as
\begin{align}
K(k,t)=\frac{m}{4\pi \rho k_B T k (k+1)}\big<R_{k}^*(0)R_{k}(t)\big>
\label{kernel}
\end{align}
and the so-called ``random force'' $R_k(t)$ is given by
\begin{align}
&R_k(t)=e^{i(1-\mathcal{P})\mathcal{L}t}R_k(0),
\label{random_force}
\end{align}
with  $\mathcal P$ the projection operator onto $A_k(0)$, $i\mathcal L$ the Liouville operator, and
\begin{align}
\label{eq_random_force}
&R_{k}(0)=\sqrt{2k+1} \bigg[\delta\ddot{\rho}_{k,0}(0)+\frac{4\pi k_B T}{m V }\frac{k(k+1)}{S(k)}\delta \rho_{k,0}(0)\bigg].
\end{align}

\subsection{Mode-coupling approximation}
The above generalized Langevin equation is exact but purely formal. The MCT of glass-forming liquids consists of 
the following approximations: (i)  One only retains in the dynamics of the random force the contribution of bilinear combinations of the density fluctuations; (ii) the operator $e^{i(1-\mathcal{P})\mathcal{L}t}$ is replaced by
$\mathcal{P}_2e^{i\mathcal{L}t}\mathcal{P}_2$  where $\mathcal{P}_2$ is a projection operator defined for any variable $X$ as
\begin{equation}
 \mathcal{P}_2 X=\sum_{\mathbf{k_1}\neq \mathbf{k_2}^*}\sum_{\mathbf{k_3}\neq \mathbf{k_4}^*} 
 B_{\mathbf{k_1}\mathbf{k_2}}\left [\langle B^*B\rangle^{-1}\right]_{\mathbf{k_1}\mathbf{k_2},\mathbf{k_3}\mathbf{k_4}}\langle B^*_{\mathbf{k_3}\mathbf{k_4}}X\rangle
 \label{projector_P2}
\end{equation}
where the vector $B$ has components $B_{\mathbf{k_1}\mathbf{k_2}}=\delta \rho_{\mathbf{k_1}}\delta \rho_{\mathbf{k_2}}$ (we have simplified the notations by writing $\mathbf{k}=(k,l)$, where $k\geq 0$, $|l|\leq k$, and its ``conjugate'' $\mathbf{k}^*=(k,-l)$. ). (iii) Finally,  the four-point density correlation function is factorized into a product of two-point correlation functions. (Note that in principle the output also involves the static triplet correlation function but the latter is almost always considered in a factorized approximation.)

After some lengthy but straightforward algebraic manipulations that are detailed in Appendix \ref{MCT_approx},  we have obtained the following expression for the mode-coupling memory function on $S^2$:
\begin{align}
&K(k,t)=\frac{\pi\rho k_B T}{2m V^2 k(k+1)} \times \nonumber\\
&\sum_{k_1\geq 0}\sum_{k_2=|k-k_1|}^{k+k_1}{Z_k(k_1,k_2)f(k_1,t)f(k_2,t)}
\label{memory_S2}
\end{align}
with 
\begin{align}
&Z_k(k_1,k_2)=(2k_1+1)(2k_2+1)\begin{pmatrix}
k_1&k_2&k\cr
0&0&0\
\end{pmatrix}^2 \times \nonumber\\
&\bigg[\big(k(k+1)+k_1(k_1+1)-k_2(k_2+1)\big)c(k_1)+\nonumber\\
&\big(k(k+1)+k_2(k_2+1)-k_1(k_1+1)\big)c(k_2)\bigg]^2S(k_1)S(k_2).
\label{vertex_S2}
\end{align}
In the above expression $\begin{pmatrix}
k_1&k_2&k\cr
0&0&0\
\end{pmatrix}$ denotes a Wigner $3j$ symbol and $c(k)$ is the direct correlation function which is  related to the static structure factor $S(k)$ by the Ornstein-Zernike equation $1/S(k)=1-\rho c(k)$. Note that the Ornstein-Zernike equation takes the same form as in the Euclidean case thanks to the appropriate choice of the Fourier transform in $S^2$.

An equation can also be derived for the self-intermediate scattering function $F_s(k,t)$ \cite{Gotze2008}, which is more conveniently computed in simulations. However, it is known that for wavenumbers corresponding to typical interatomic distances the temperature dependences of the relaxation times associated with the dynamic structure factor and the self-intermediate scattering function are similar. More importantly for our purpose here, the critical temperatures $T_c$ at which the dynamics are predicted to freeze are identical.

\subsection{Euclidean limit}
In the two-dimensional euclidean space the MCT equations have already been studied \cite{Bayer2007,Weysser2011} and read
\begin{align}
&\ddot f(k,t)+\frac{k_B T}{m}\frac{k^2}{S(k)}f(k,t)+\int_{0}^t{dt' K(k,t-t')\dot f(k,t')}=0
\label{MCT_euclidean}
\end{align}
with 
\begin{align}
 K(k,t)=&\frac{\rho k_B T}{8\pi^2 m}
\int_{\mathbb{R}^2} d\vec{k}_1|\tilde{V}_{\vec{k_1},\vec{k}-\vec{k_1}}|^2S(k_1)S(|\vec{k}-\vec{k_1}|)\times 
 \nonumber\\
 &f(k_1,t) f(|\vec{k}_1-\vec{k}|,t)
\end{align}
and
\begin{align}
&\tilde{V}_{\vec{k_1},\vec{k}-\vec{k_1}}=\frac{\vec{k}.\vec{k_1}}{k}c(k_1)+\frac{\vec{k}.(\vec{k}-\vec{k_1})}{k}c(|\vec{k}-\vec{k_1}|).
\end{align}
To facilitate the comparison of the above expressions with the limit of the spherical case derived in the previous subsection when $R\rightarrow \infty$, we rewrite the function $K(k,t)$ as
\begin{eqnarray}
&K(k,t)=\frac{\rho k_B T}{8\pi^2k^2m}\int_{0}^{\infty}dk_1 k_1\int_{|k-k_1|}^{k+k_1}dk_2 k_2 V_{k_1,k_2}^2 \nonumber\\
& \times \frac{S(k_1)S(k_2)f(k_1,t)f(k_2,t)}{\sqrt{-(k^4+k_1^4+k_2^4)+2(k^2k_1^2+k^2k_2^2+k_1^2k_2^2)}}
\label{kernel_euclidean}
\end{eqnarray}
with a new vertex $V_{k_1,k_2}$ defined as
\begin{equation}
\label{vertex_new}
 V_{k_1,k_2}=(k^2-k_2^2+k_1^2)c(k_1)+(k^2+k_2^2-k_1^2)c(k_2).
\end{equation}
Note that Eqs. (\ref{kernel_euclidean}) and (\ref{vertex_new}) only depend on the modulus  of the wavevectors.

When $R\rightarrow \infty$ in Eqs. (\ref{exact}), (\ref{memory_S2}) and (\ref{vertex_S2}), the wavenumbers $k,k_1,\cdots$, which all go to infinity, must be rescaled by $R$, \textit{i.e.}, $k=R \hat k$ with $\hat k \in \mathbb{R}^+$. Under such a rescaling, the structure factor on $S^2$, $S(k)$ converges to $S(\hat k)$, where the latter is now the Euclidean structure factor, and similarly $F(k,t)$ converges to its Euclidean counterpart $F(\hat k,t)$. The coefficient of the term in $f(k,t)$ in Eq. (\ref{exact}), which is like the square of an effective frequency, goes to $k_B T \hat k^2/[mS(\hat k)]$, which is precisely the coefficient in the Euclidean equation, Eq. (\ref{MCT_euclidean}).

To treat the memory term, one must use the large-wavenumber limit of the Wigner $3j$ symbols, \cite{Borodin1978}
\begin{align}
&\begin{pmatrix}
k_1&k_2&k_3\cr
0&0&0\
\end{pmatrix}^2\nonumber\\
&\sim \frac{2}{\pi\sqrt{-(k_1^4+k_2^4+k_3^4)+2(k_1^2k_2^2+k_1^2k_3^2+k_2^2k_3^2)}},
\label{wigner_limit}
\end{align}
where $|k_1-k_2|\leq k_3\leq k_1+k_2$  and $k_1$, $k_2$, $k_3\rightarrow\infty$. The $3j$ symbols in Eq.~(\ref{wigner_limit}) are only nonzero when
the sum $k_1+k_2+k_3$ is even.

After introducing the above property in Eq. (\ref{vertex_S2}), using the rescaling of the wavenumbers, and replacing the sums in Eq. (\ref{memory_S2}) by integrals, \textit{i.e.}, 
\begin{align}
& \sum_{k_1\geq 0}\sum_{k_2=|k-k_1|}^{k+k_1}\begin{pmatrix}
k_1&k_2&k_3\cr
0&0&0\
\end{pmatrix}^2 \cdots \sim \frac{1}{2}\int_{0}^\infty d\hat{k}_1\int_{|\hat{k}-\hat{k}_1|}^{\hat{k}+\hat{k}_1}d\hat{k}_2 \nonumber\\
&\frac{2}{\pi \sqrt{-(\hat{k}_1^4+\hat{k}_2^4+\hat{k}^4)+2(\hat{k}_1^2\hat{k}_2^2+\hat{k}_1^2\hat{k}^2+\hat{k}_2^2\hat{k}^2)}} \cdots \,,
\end{align}
one arrives at Eqs. (\ref{kernel_euclidean},\ref{vertex_new}). Note that there is a factor $1/2$ in front of the integrals because of 
the condition that $k_1+k_2+k_3$ is even. This completes the proof that the MCT equations on $S^2$, which we have derived above,  
converge to the MCT equations of the Euclidean plane when the radius $R$ of the sphere goes to infinity.

\section{Mode-coupling singularity and comparison with simulation data}

\subsection{Predicted dynamical singularity}
As mentioned in the Introduction, our goal is two-fold. On the one hand, we would like to study whether curvature induces a slowdown of relaxation whose nature is qualitatively different from that of standard glass-forming liquids. Such a qualitative difference could be signaled by a dynamical mode-coupling transition of a different character than the discontinuous one predicted for glass-forming liquids in 2- and 3-dimensional Euclidean spaces. On the other hand, increasing curvature is found to speed up the dynamics and we want to check if, and to which extent, the MCT is able to capture this trend. For studying these questions, it is sufficient to consider the nature and the curvature dependence of the dynamical singularity of the MCT equations.

We therefore consider  the long-time limit $f(k)=\lim_{t\rightarrow\infty}f(k,t)$ of the MCT equation, which is known as the nonergodicity parameter. 
Within the MCT framework \cite{Gotze1988,Gotze2008}, a dynamical transition at some critical $T_c$ separates a high-temperature, ergodic phase, in which $f(k)$ is equal to zero, from a low-temperature,
nonergodic phase in which $f(k)>0$. For glass-forming liquids in Euclidean space, the  transition is discontinuous, with  a jump of $f(k)$ at $T_c$ (a B-type singularity), whereas in other situations, such as liquids confined in disordered environments, the transition may be continuous (an A-type singularity) \cite{Krakoviack2005}.

The equation for the nonergodicity
parameter is easily obtained from the full MCT equation in the long-time limit by taking the time Laplace transform, and one obtains that \cite{Gotze2008}
\begin{equation}
\label{nonergodicity_factor}
\frac{f(k)}{1-f(k)}=M_\infty(k),
\end{equation}
where, for a liquid on $S^2$,
\begin{equation}
\label{kernel_S2}
M_\infty(k)=\frac{\rho S(k)}{8Vk^2(k+1)^2}\sum_{k_1,k_2}{Z_k(k_1,k_2)f(k_1)f(k_2)}
\end{equation}
with $Z_k(k_1,k_2)$ given in Eq.~(\ref{vertex_S2}).

A numerical solution of Eqs. (\ref{nonergodicity_factor},\ref{kernel_S2}) can be obtained with an iterative procedure,
\begin{equation*}
\frac{f^{(i+1)}(k)}{1-f^{(i+1)}(k)}=M_\infty^{(i)}(k),
\end{equation*}
where $M_\infty^{(i)}(k)$ is calculated by using the nonergodicity parameter of the previous iteration $\{f^{(i)}(k')\}$. 
For the sake of simplicity, the initial condition is chosen as $f^{(0)}(k)=1$. 
The iterative procedure is stopped when  the  condition $\sum_{k}|f^{(i+1)}(k)-f^{(i)}(k)|^2<10^{-8}$ is fulfilled. The convergence 
is usually obtained after few iterations except close to the transition where the number of iterations is of order $100$. 

For the Euclidean model, the memory term is given by
\begin{eqnarray}
\label{kernel_E2}
&M_\infty(k)=\frac{\rho S(k)}{8\pi^2k^4}\int_{0}^{\infty}dk_1 k_1\int_{|k-k_1|}^{k+k_1}dk_2 k_2 V_{k_1,k_2}^2 \nonumber\\
&\frac{S(k_1)S(k_2)f(k_1)f(k_2)}{\sqrt{-(k^4+k_1^4+k_2^4)+2(k^2k_1^2+k^2k_2^2+k_1^2k_2^2)}}\,.
\end{eqnarray}

The numerical integration is a little more subtle, because at a given $k_1$ the integrand has singularities at the  bounds of the integral over $k_2$.  By using a trapezoidal rule, the numerical method goes to the exact result as the inverse of the square root of the number of points, which cannot provide a sufficient accuracy for the iterative method. To improve the computation of the integral, we have estimated it by using two different steps for the integration, and it is then easy to show that one can obtain a convergence to the exact value as the number of chosen points to the power $3/2$.

\begin{figure}[t!]
\begin{center}   
  \includegraphics[width=0.45\textwidth]{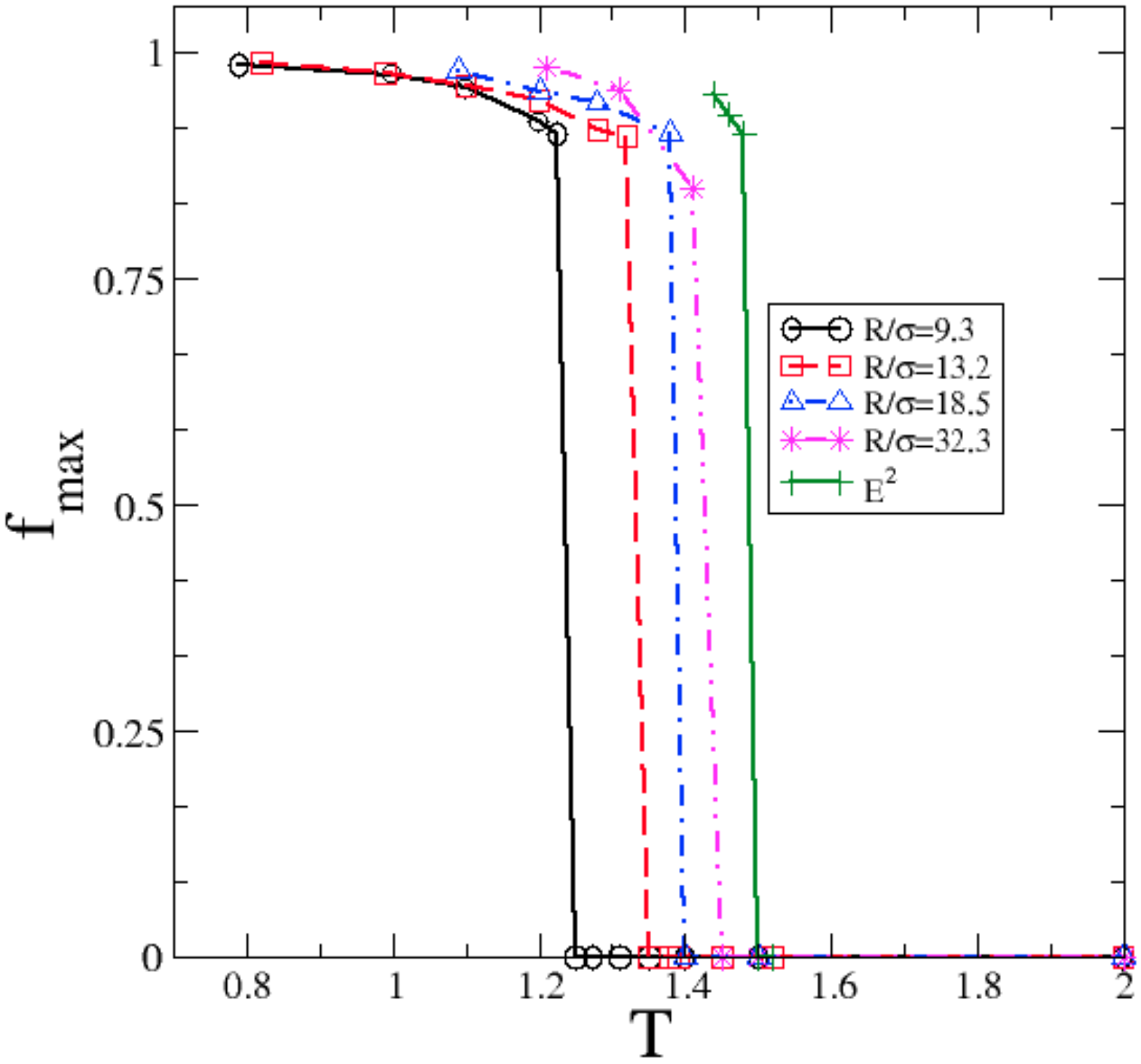}
   \caption{MCT prediction for the temperature dependence of the nonergodicity parameter $f_{\text{max}}=f(k_{\text{max}})$ 
   for $S^2$ with different values of $R/\sigma$ and for the Euclidean case. For the largest sphere, $R/\sigma=32.3$, the numerical uncertainty is larger
   which explains that  the magnitude of the discontinuity at $T_c$ is slightly too small.}
   \label{fk_fonc_T}
\end{center}
\end{figure}

The solution of Eqs. (\ref{nonergodicity_factor}), (\ref{kernel_S2}), and (\ref{kernel_E2}) is shown in Fig. \ref{fk_fonc_T} where we plot the temperature dependence of the nonergodicity parameter $f_\text{max}=f(k_{\text{max}})$, where $k_{\text{max}}$ corresponds to the value of the first peak of $S(k)$, for several values of the reduced radius of curvature and for the Euclidean case. Note that in the latter case, the predicted dynamical transition takes place at a temperature that is higher than the ordering transition in the plane so that the associated static structure factors are those of a liquid. (We have varied the system size and checked that the finite-size effects are negligible for the systems under study.)  

\begin{figure}[h!]
\begin{center}   
 \includegraphics[width=0.45\textwidth]{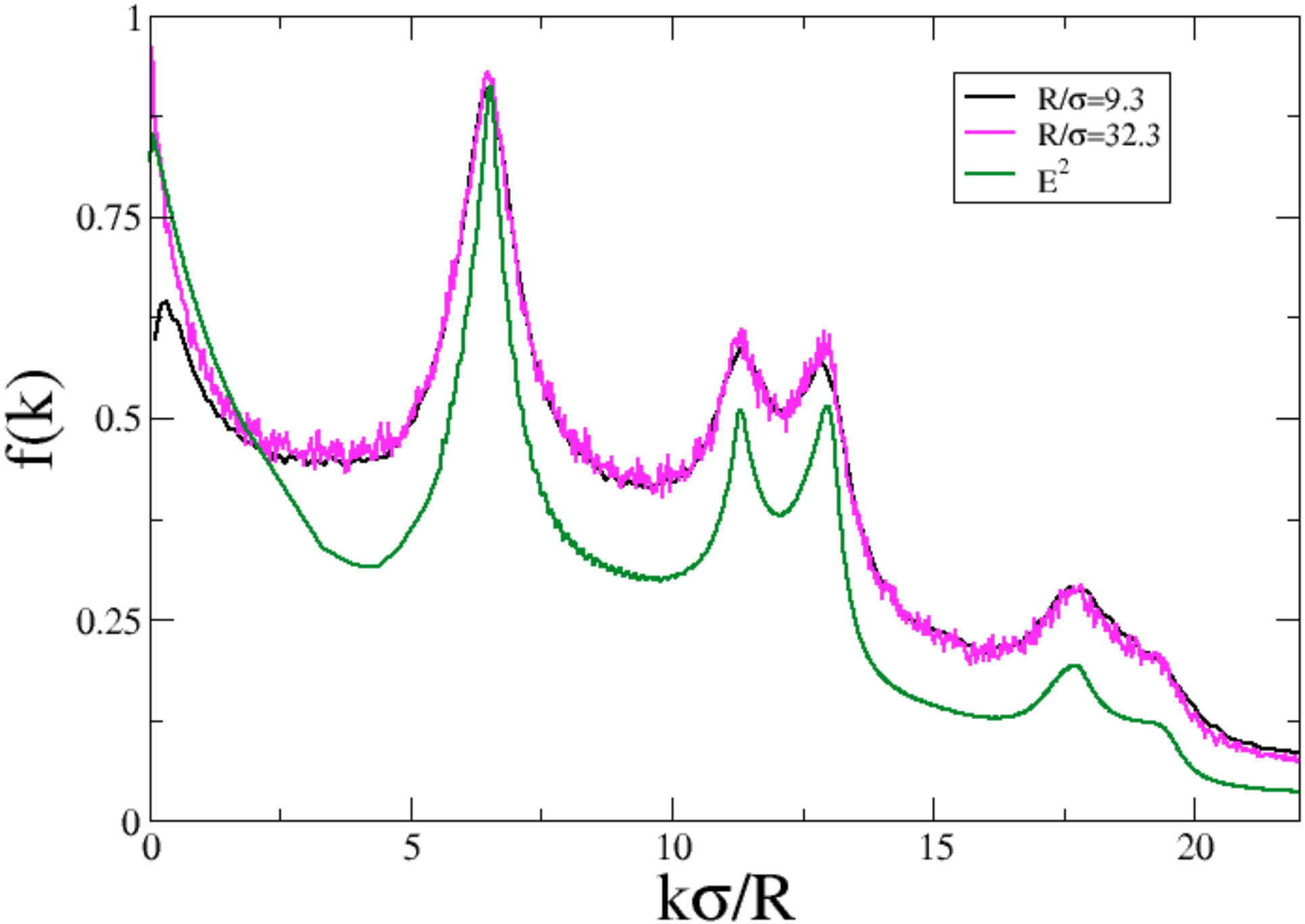}
   \caption{MCT prediction of the nonergodicity parameter $f(k)$ versus the reduced wave number ($k\sigma/R$ for $S^2$ and $k\sigma$ for $E^2$) at the transition $T=T_c^{-}$, for $S^2$ with $R/\sigma=9.3$ and $32.3$, and for the Euclidean case.}
   \label{fk_Tcomp}
 \end{center}
\end{figure}

The first observation is that the transition is discontinuous in $S^2$ just as it is in Euclidean space. Therefore, from the MCT perspective, there is nothing anomalous in the glass transition of liquids in curved space. The second observation is that the predicted critical temperature $T_c( R)$ decreases with increasing curvature, from $1.49$ in the Euclidean plane to $1.24$ for the largest curvature corresponding to $R/\sigma=9.3$. In Fig. \ref{fk_Tcomp} we also illustrate the $k$-dependence of the nonergodicity parameter at the transition $T_c$ for the Euclidean plane and one curvature.

From the above MCT equations, one can derive some analytical results. First, as the dynamical singularity in $S^2$ is of the same B-type as the singularity found for glass-forming liquids in Euclidean space, one finds the same critical behavior with in particular a square root dependence of the nonergodicity parameter in the glass phase,
\begin{equation}
f(k,T)-f(k,T_c) \propto \left (\frac{T-T_c}{T_c}\right )^{1/2}
\end{equation}
when $T\rightarrow T_c^{-}$.

\begin{figure}[h!]
\begin{center}   
   \includegraphics[width=0.45\textwidth]{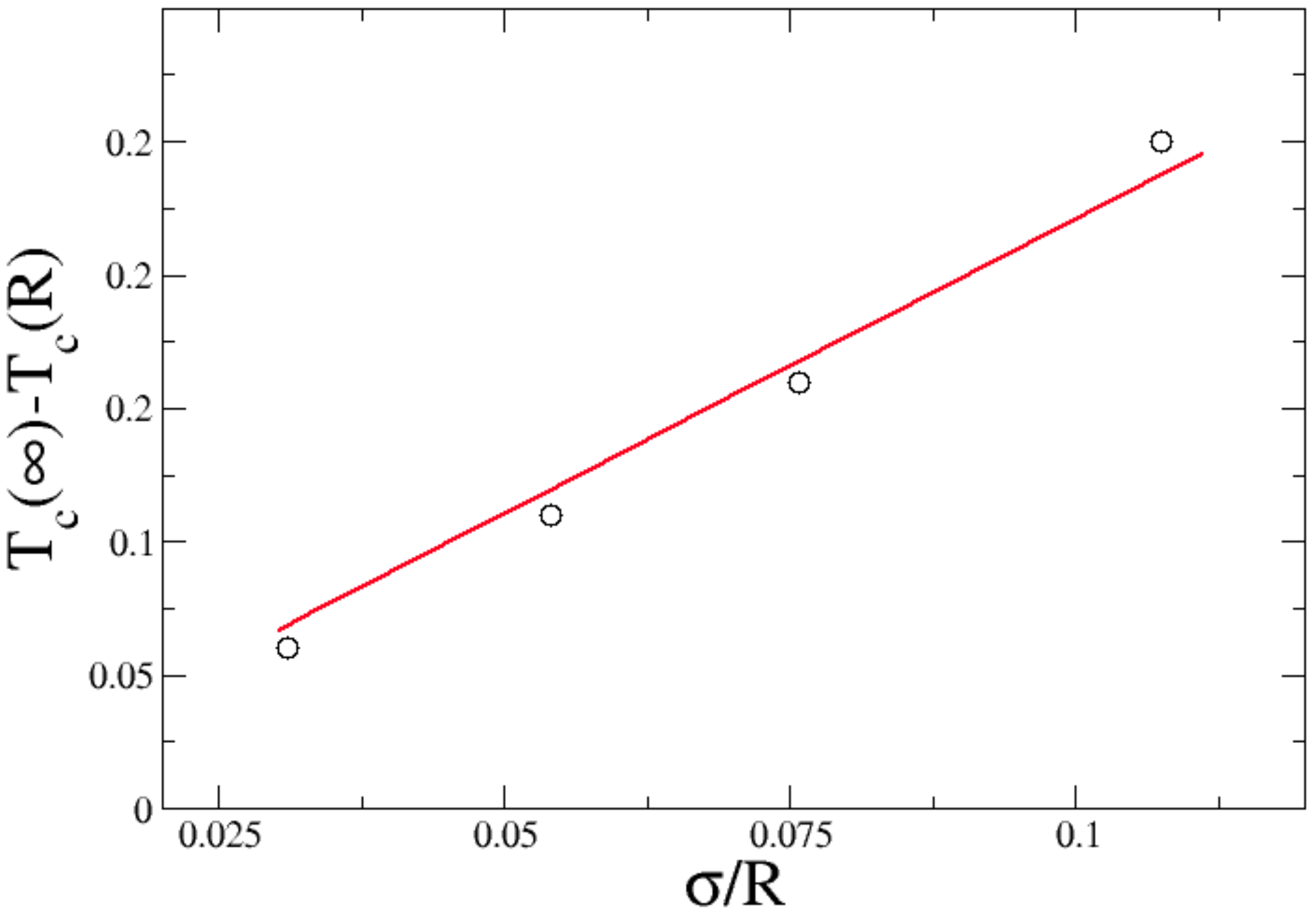}
   \caption{Curvature dependence of the predicted MCT critical temperature: $T_c(\infty)-T_c( R)$ versus $\sigma/R$ for $R/\sigma=9.3$, $13.2$, $18.5$ and $32.3$. The value of the transition temperature on the Euclidean plane is $T_c(\infty)\simeq 1.49$. The full line is the expected linear asymptotic behavior.}
   \label{fig:Tc_fonc_R}
\end{center}
\end{figure}

In addition, one can study the asymptotic behavior of the MCT equations when $R\rightarrow \infty$. We find that the first correction to the kernel $K(k,t)$ is in $\sigma/R$, which leads to an expected behavior of the transition temperature as
\begin{equation}
T_c(\infty)-T_c( R) \sim \frac{\sigma}{R}\,
\end{equation}
when approaching the flat (Euclidean) space limit. We plot in Fig. \ref{fig:Tc_fonc_R}  $T_c(\infty)-T_c( R)$ versus $\sigma/R$. Within the numerical accuracy the results are compatible with a linear behavior.


\subsection{Comparison with simulation data}
We would like to compare the predictions of the MCT with the simulation data. This is not an easy task as the dynamical transition predicted by the MCT is not observed in finite-dimensional systems and, as a result, the predictions are quantitatively inaccurate. The conventional procedure to compare theory and simulation is to try to empirically fit the numerical data to the theoretical results by letting the temperature of the postulated dynamical singularity be an adjustable parameter. This empirical temperature, $T_c^{(emp)}$, can be further used as a rescaling parameter to improve the quality of the comparison. Our interest here is not in pursuing such a rescaling, but the empirically determined $T_c^{(emp)}$ is a useful quantity to compare predicted and observed trends in a restricted domain of relaxation times.

From the simulation data for the intermediate scattering function we have extracted an estimate of the relaxation time $\tau$. We have then fitted the temperature dependence of the latter to the MCT prediction by using the scaling law $\tau \propto (T-T_c)^{-\gamma}$ over an ``optimal'' range of relaxation time. To restrict the number of adjustable parameters, we have set $\gamma = 2.38$ according to results obtained in the two-dimensional Euclidean space \cite{Bayer2007}. $T_c\equiv T_c^{(emp)}$ is then obtained from the best fit to the data. 

\begin{figure}[h!]
\begin{center}   
 \includegraphics[width=0.5\textwidth]{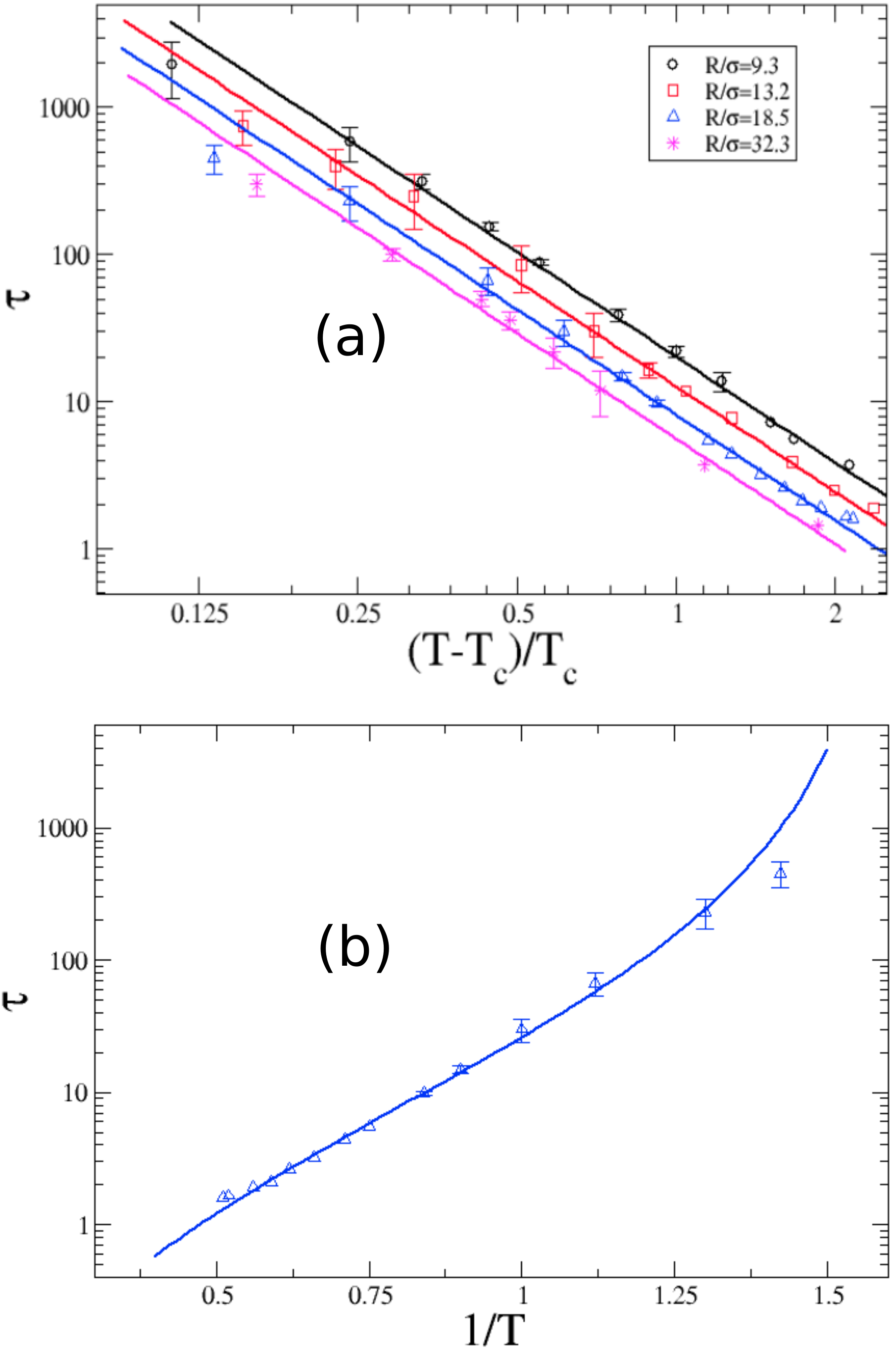}
   \caption{a) Log-log plot of the relaxation time $\tau$ versus $(T-T_c^{(emp)})/T_c^{(emp)}$ for different curvatures. $T_c^{(emp)}$ is chosen to obtain the best collapse of the simulation data to the scaling law $\tau \propto (T-T_c)^{-\gamma}$ with the value of the exponent fixed to $\gamma=2.38$. b) Illustration of the same MCT fit (full line) on an Arrhenius plot of $\log(\tau)$ versus $1/T$ for $R/\sigma=18.5$.}
   \label{tau_MCT_fit}
\end{center}
\end{figure}

Fig. \ref{tau_MCT_fit}a displays a log-log plot of the simulation data with the associated MCT fits for the 4 different curvatures studied. (One observes that the prefactor of the scaling law  depends on the reduced curvature and decreases with $R/\sigma$ but the physical significance of this trend is unclear.) In addition, Fig. \ref{tau_MCT_fit}b illustrates the same MCT fit now shown on an Arrhenius plot for one curvature. As well known, the quality of the fit is rather poor and is only valid over at most a couple of orders of magnitude in $\tau$ \cite{Berthier2010b}. Nothing here is specific of curved space.

The predicted and empirically determined critical temperatures are given in Table 1 for the 4 different curvatures under study. Several comments can be made:

1) As already found in all other MCT studies of glass-forming liquids, the theoretically determined $T_c$ is much higher than the empirically determined one, with a factor 2 or more between them.

2) The theory predicts the correct trend with increasing curvature, \textit{i.e.}, that the slowdown becomes weaker as curvature increases: this is signaled by a decrease of $T_c( R)$ with decreasing $R/\sigma$.

3) The magnitude of the variation with curvature is dramatically underestimated by the theory: the reduction in $T_c$ between the lowest nonzero curvature ($R/\sigma= 32.3$) and the highest one ($R/\sigma= 9.3$) is about 4 times too small in the theory. The qualitative trend is right but the quantitative estimate of the effect is completely off.

\begin{table}[b]
\caption{\label{tab:table1}%
Table of temperatures
}
\begin{ruledtabular}
\begin{tabular}{lccr}
\textrm{$R/ \sigma$}&
\textrm{$T_c$}&
\textrm{$T_c^{(emp)}$}&
\textrm{$T_c/T_c^{(emp)}$}
 \\
 \hline
 $9.3$ &$1.24$&$0.45$ &$2.8$\\
 $13.2$ &$1.33$& $0.53$& $2.7$\\
 $18.5$ & $1.38$& $0.62$ & $2.5$\\
  $32.3$ & $1.43$ & $0.70$& $2.2$
\end{tabular}
\end{ruledtabular}
\end{table}

\section{Conclusion}

We have studied the dynamics of a one-component liquid constrained on a spherical substrate and investigated how the mode-coupling theory (MCT) can describe the new features brought by the presence of curvature. To this end we have derived the MCT equations for the time-dependent pair correlation function 
of the density fluctuations in a spherical geometry. The ratio of the radius of the embedding $2$-sphere over the atomic diameter is a new control parameter. 

We find that, as seen from the MCT, the slow dynamics of liquids in curved space at low temperature does not qualitatively differ from that of glass-forming
liquids in Euclidean space 
\footnote{This is valid at least in 2 dimensions: for a discussion of the differences between liquids in $2$- and $3$-dimensional (Euclidean) spaces, 
see \cite{Flenner2015}.}. The dynamical transition is of the discontinuous (B) type in all cases. The MCT predicts the correct trend for the evolution of the relaxation slowdown with curvature but is dramatically off at a quantitative level. As found in other cases \cite{Berthier2010b}, the MCT is unable to describe situations where significant differences in the temperature dependence of the dynamics come with only modest changes in the static pair structure.

\begin{acknowledgements}
The authors gratefully acknowledge V. Krakoviack for useful discussions.
\end{acknowledgements}

\appendix

\section{Fourier transform on $S^2$}
\label{Fourier_S2}

The sphere is a manifold with a finite area, so that the Fourier transform of a function of the angular coordinates, $f(\Omega)$,
is given by a discrete number 
\begin{align}
f_{k,l}=&\sqrt{\frac{4\pi}{2k+1}}R^2\int_{S^2}{d\Omega  Y_{k,l}^*(\Omega)f(\Omega)},\label{eq:dft}
\end{align}
where $k$ and $l$ are  integers such that $k\geq 0$ and  $-k\leq l\leq k$, and $Y_{k,l}^*(\Omega)$ denotes the complex conjugate of the spherical harmonics $Y_{k,l}(\Omega)$ which is defined as
\begin{equation}
Y_{k,l}(\Omega)=(-1)^l\sqrt{\frac{2k+1}{4\pi}\frac{(k-l)!}{(k+l)!}}\, P_{k,l}(\cos\theta)e^{i l\phi},
\end{equation}
where $P_{k,l}(\cos\theta)$ is an associated Legendre function of first kind. With this definition, $Y_{0,0}=1/\sqrt{4\pi}$ and $\int_{S^2}{d\Omega  Y_{k,l}^*(\Omega)Y_{k',l'}(\Omega)}=\delta_{k,k'}\delta_{l,l'}$.

The corresponding inverse Fourier transform is then
\begin{align}
f(\Omega)=&\sqrt{\frac{2k+1}{4\pi}}\frac{1}{R^2}\sum_{k\geq 0}\sum_{l=-k}^k{Y_{k,l}(\Omega)f_{k,l} }.
\end{align}

For an isotropic function $f(\Omega)=f(\theta)$,  $f_{k,l}=f_{k,0}\equiv f(k)$. In the limit $R\rightarrow \infty$, after having introduced $\hat k=k/R$ and $\tilde{f}(r)=f(r/R)$,   one obtains that  $f(k)$ goes to
\begin{equation}
\tilde{f}(\hat k)=2\pi\int_{0}^\infty dr r J_0(\hat kr)\tilde{f}( r)\,,
\end{equation}
which is the usual Fourier transform in 2 dimensions.
A similar reasoning holds for recovering the inverse Fourier transform.

\section{Mode-coupling approximation on $S^2$ \label{MCT_approx}}

We summarize here the main steps of the derivation of the MCT equations on $S^2$.

The projection operator $\mathcal{P}_2$ on the pair products of density modes, when applied to the random force, leads to
\begin{equation}
\mathcal{P}_2 R_k=\sum_{\mathbf{k_1}\neq \mathbf{k_2}^*} V_k(\mathbf{k_1},\mathbf{k_2}) B_{\mathbf{k_1}\mathbf{k_2}}, 
\end{equation}
with
\begin{equation}
V_{k}(\mathbf{k_1},\mathbf{k_2})=\sum_{\mathbf{k_3}\neq\mathbf{k_4}^*}{\big<\delta \rho^*_{\mathbf{k_3}}\delta \rho_{\mathbf{k_4}}^* R_{k}\big>[D^{-1}]_{\mathbf{k_1}\mathbf{k_2},\mathbf{k_3}\mathbf{k_4}}},
\label{Vk}
\end{equation}
where $[D]_{\mathbf{k_1}\mathbf{k_2},\mathbf{k_3}\mathbf{k_4}}=\big<B^*_{\mathbf{k_1}\mathbf{k_2}}B_{\mathbf{k_3}\mathbf{k_4}}\big>$ 
and we recall the notation $\mathbf k_i\equiv (k_i,l_i)$ and $\mathbf k^*_i\equiv (k_i,-l_i)$. 

Within the Gaussian approximation, the 4-point density correlation function $[D]_{\mathbf{k_1}\mathbf{k_2},\mathbf{k_3}\mathbf{k_4}}$,  with $\mathbf{k_1}\neq \mathbf{k_2}^*$ and $\mathbf{k_3}\neq \mathbf{k_4}^*$, is given by
\begin{eqnarray}
[D]_{\mathbf{k_1}\mathbf{k_2},\mathbf{k_3}\mathbf{k_4}}&\approx&\big<\delta \rho_{\mathbf{k_1}}^*\delta\rho_{\mathbf{k_3}}\big>\big<\delta \rho_{\mathbf{k_2}}^*\delta\rho_{\mathbf{k_4}}\big>\nonumber\\
&&+\big<\delta \rho_{\mathbf{k_1}}^*\delta\rho_{\mathbf{k_4}}\big>\big<\delta \rho_{\mathbf{k_2}}^*\delta\rho_{\mathbf{k_3}}\big>\nonumber\\
&=&(\delta_{\mathbf{k_1},\mathbf{k_3}}\delta_{\mathbf{k_2},\mathbf{k_4}}+\delta_{\mathbf{k_1},\mathbf{k_4}}\delta_{\mathbf{k_2},\mathbf{k_3}})\nonumber\\
&&\times N^2 \frac{S(k_1)S(k_2)}{(2k_1+1)(2k_2+1)} \,,
\label{denom}
\end{eqnarray}
so that
\begin{equation}
\begin{aligned}
&[D^{-1}]_{\mathbf{k_1}\mathbf{k_2},\mathbf{k_3}\mathbf{k_4}}= \\&
\frac{(2k_1+1)(2k_2+1)}{4N^2S(k_1)S(k_2)}\big(\delta_{\mathbf{k_1},\mathbf{k_3}}\delta_{\mathbf{k_2},\mathbf{k_4}}+\delta_{\mathbf{k_1},\mathbf{k_4}}\delta_{\mathbf{k_2},\mathbf{k_3}}\big).
\end{aligned}
\end{equation}

The MCT memory function is obtained from
\begin{equation}
K(k,t)=\frac{m}{4\pi \rho k_B T k (k+1)}\big<[e^{i\mathcal L t}\mathcal{P}_2 R_{k}][\mathcal{P}_2 R_{k}]^*\big>\,,
\end{equation}
which, by using the above expressions and, once more, the Gaussian approximation, can be rewritten as
\begin{equation}
K(k,t)=\frac{m}{2\pi \rho k_B T k (k+1)} \sum_{\mathbf{k_1}\neq \mathbf{k_2}^*} \vert V_k(\mathbf{k_1},\mathbf{k_2})\vert^2 F(k_1,t)F(k_2,t) \,.
\label{eq_memory_gaussian}
\end{equation}

Given the definition of the random force in Eq. (\ref{eq_random_force}), the calculation of $\big<\delta \rho^*_{\mathbf{k_3}}\delta \rho_{\mathbf{k_4}}^* R_{k}\big>$ appearing in Eq. (\ref{Vk}) 
requires the consideration of two terms:
\begin{eqnarray}
U_{\mathbf{k_3},\mathbf{k_4},\mathbf{k}}&=&\big<\delta \dot{\rho}_{\mathbf{k_3}}^*\delta \rho_{\mathbf{k_4}}^*\delta\dot{\rho}_{\mathbf k}\big>,\\
T_{\mathbf{k_3},\mathbf{k_4},\mathbf{k}}&=&\big<\delta \rho_{\mathbf{k_3}}^*\delta \rho_{\mathbf{k_4}}^*\delta\rho_{\mathbf k}\big>,
\end{eqnarray}
where, here and below, $\mathbf k=(k,0)$.

The $3$-point density correlation function $T_{\mathbf{k_3},\mathbf{k_4},\mathbf{k}}$ is expressed in the convolution approximation as
\begin{equation}
T_{\mathbf{k_3},\mathbf{k_4},\mathbf{k}}\approx  N\begin{pmatrix}
k&k_3&k_4\cr
0&l_3&l_4\
\end{pmatrix}
\begin{pmatrix}
k&k_3&k_4\cr
0&0&0\
\end{pmatrix}S(k)S(k_3)S(k_4).
\end{equation}

The other correlator $U_{\mathbf{k_3},\mathbf{k_4},\mathbf{k}}$ can be rewritten by using the definition of the Fourier transform for the density (see Appendix \ref{Fourier_S2}) and the following expression for its derivative:
\begin{equation}
\delta \dot{\rho}_{\mathbf{k}}=\sqrt{\frac{4\pi}{2k+1}}\sum_{i=1}^N  {\bf v}_i.\nabla Y^*_{\mathbf{k}}.
\end{equation}
$U_{\mathbf{k_3},\mathbf{k_4},\mathbf{k}}$ is then expressed in terms of sums over $i$ and $j$ of $\big<\nabla Y_{\mathbf{k_3}}(\Omega_i)\nabla Y_{\mathbf{k}}(\Omega_i)Y_{\mathbf{k_4}}(\Omega_j)\big>$. This average is easily calculated thanks to the three following properties:

1) A product of gradients is expressed with only Laplacians through
\begin{equation}
\nabla f.\nabla g=\frac{1}{2}[\Delta(fg)-f\Delta g -g\Delta f ].
\end{equation}

2) The spherical harmonics are Eigenfunctions of the Laplacian: $\Delta Y_{k,l}=-k(k+1)Y_{k,l}$.

3) A product of two spherical harmonics is a linear combination of spherical harmonics with coefficients depending on $3j$ symbols:
\begin{align}
Y_{k,l}Y_{k_3,l_3}&=\sum_{k',l'}\sqrt{\frac{(2k+1)(2k_3+1)(2k'+1)}{4\pi}}\nonumber\\
&\times \begin{pmatrix}
k_3&k'&k\cr
l_3&l'&l\
\end{pmatrix}
\begin{pmatrix}
k_3&k'&k\cr
0&0&0\
\end{pmatrix}Y_{k',l'}^*.
\end{align}

One then obtains
\begin{eqnarray}
U_{\mathbf{k_3},\mathbf{k_4},\mathbf{k}}&=&\frac{Nk_B T}{mV}2\pi [k(k+1)+k_3(k_3+1)-k_4(k_4+1)] \nonumber \\
&& \times \begin{pmatrix}
k_3&k_4&k\cr
l_3&l_4&0\
\end{pmatrix}
\begin{pmatrix}
k_3&k_4&k\cr
0&0&0\
\end{pmatrix}
S(k_4). 
\end{eqnarray}

Summing up all the preceding results and introducing the direct correlation function $c(k)=(1/\rho)[1-1/S(k)]$ then lead to
\begin{equation}
\begin{aligned}
&V_k(\mathbf{k_1},\mathbf{k_2})=\frac{\pi \rho k_B T}{mV} \sqrt{(2k+1)(2k_1+1)(2k_2+1)} \times \\&
\begin{pmatrix}
k_1&k_2&k\cr
l_1&l_2&0\
\end{pmatrix}
\begin{pmatrix}
k_1&k_2&k\cr
0&0&0\
\end{pmatrix}
\Big ( [k(k+1)+ k_1(k_1+1)-  \\& 
k_2(k_2+1)]c(k_1)+
[k(k+1)+k_2(k_2+1)-k_1(k_1+1)]c(k_2) \Big ) . 
\end{aligned}
\end{equation}
After inserting the above result in Eq. (\ref{eq_memory_gaussian}), one readily finds the expressions in Eqs. (\ref{memory_S2},\ref{vertex_S2}) of the main text.


%
\end{document}